# Conceptual design of 20 T hybrid accelerator dipole magnets

P. Ferracin, G. Ambrosio, M. Anerella, D. Arbelaez, L. Brouwer, E. Barzi, L. Cooley, J. Cozzolino, L. Garcia Fajardo, R. Gupta, M. Juchno, V.V. Kashikhin, F. Kurian, V. Marinozzi, I. Novitski, E. Rochepault, J. Stern, G. Vallone, B. Yahia, A.V. Zlobin

*Abstract*— Hybrid magnets are currently under consideration as an economically viable option towards 20 T dipole magnets for next generation of particle accelerators. In these magnets, High Temperature Superconducting (HTS) materials are used in the high field part of the coil with so-called "insert coils", and Low Temperature Superconductors (LTS) like Nb$_3$Sn and Nb-Ti superconductors are used in the lower field region with so-called "outsert coils". The attractiveness of the hybrid option lays on the fact that, on the one hand, the 20 T field level is beyond the Nb$_3$Sn practical limits of 15-16 T for accelerator magnets and can be achieved only via HTS materials; on the other hand, the high cost of HTS superconductors compared to LTS superconductors makes it advantageous exploring a hybrid approach, where the HTS portion of the coil is minimized. We present in this paper an overview of different design options aimed at generating 20 T field in a 50 mm clear aperture. The coil layouts investigated include the Cos-theta design (CT), with its variations to reduce the conductor peak stress, namely the Canted Cos-theta design (CCT) and the Stress Management Cos-theta design (SMCT), and, in addition, the Block-type design (BL) including a form of stress management and the Common-Coil design (CC). Results from a magnetic and mechanical analysis are discussed, with particular focus on the comparison between the different options regarding quantity of superconducting material, field quality, conductor peak stress, and quench protection.

## I. INTRODUCTION

THE superconducting magnet community, which is working on the next generation of magnets for future particle colliders, has being considering the option of a "20 T" dipole magnet since approximately 20 years. The first proposal was formulated by P. McIntyre *et al.* [1], who, considering the nominal field of 8.3 T of the LHC dipoles, explored in 2005 the possibility of a 24 T dipole magnet for an "LHC tripler". In 2011, the design studies carried out by E. Todesco, *et al.* [2]-[3] and by R. Gupta, *et al.* [4] were focused on dipole magnets generating an operational field of 20 T, with the goal of "opening the way for a 16.5 TeV beam energy accelerator in the LHC tunnel", being 7 TeV the nominal beam energy of the LHC. A similar field level was then considered for the future Super proton-proton Collider (SppC) in China by G. Sabbi, *et al.* [5] and by Q. Xu *et al.* [6], and for the European Future Circular Collider (FCC) by J. van Nugteren, *et al.* [7].

A different viewpoint to explain the rationale behind the idea of a 20 T accelerator magnet lays in the continuous push towards high field magnets to achieve higher collision energy [8], and in particular to a sort of "4 T step" that has characterized the R&D on superconducting accelerator magnets in the last two decades. In fact, a 4 T jump has characterized the increase in field from the Nb-Ti dipole magnets installed in the LHC [9] to the Nb$_3$Sn magnets (in this case quadrupoles) planned for the HL-LHC project and expected to operate with a conductor peak field approaching 12 T [10]. The FCC design study has then worked on arc dipoles with a bore field of 16 T, a level considered as the practical limit for the Nb$_3$Sn technology [11]-[12]. In this landscape, the next natural milestone is represented by a 20 T magnet, where so called High Temperature Superconductor (HTS), in particular Bi2212 [13] and REBCO [14], need to be adopted to push the field beyond the Low Temperature Superconductors (namely Nb$_3$Sn) limits.

As a last consideration, one has to take into account the still relevant higher cost of HTS conductor compared to Nb$_3$Sn. The significant difference in superconductor price justifies investigating the hybrid option, where Nb$_3$Sn is included in the coil design to minimize the quantity of HTS material. This option was recently tested with the FRESCA2 large aperture dipole magnet as outsert and with the HTS EUCARD2 coil as insert [15]-[16], and explored in a recent conceptual design study [17].

We describe in this paper three conceptual designs of a 20 T hybrid magnet. The work is a continuation of a preliminary and broader investigation carried out in [18] as part of the US Magnet Development Program (MDP) [19]. After summarizing in Section I the design criteria, in Section II we perform a parametric analysis using sector coils. In Section III we then describe cos-theta, block and common-coil designs, focusing on magnetic parameters and coil stresses. Some consideration regarding fabrication options and challenges will also be provided.

This work was supported by the U.S. Department of Energy, Office of Science, Office of High Energy Physics, through the US Magnet Development Program (*Corresponding author: Paolo Ferracin*).

P. Ferracin, D. Arbelaez, L. Brouwer, L. Garcia Fajardo, M. Juchno and G. Vallone are with Lawrence Berkeley National Lab, Berkeley, CA 94720, USA (e-mail: pferracin@lbl.gov).
G. Ambrosio, E. Barzi, V.V. Kashikhin, V. Marinozzi, I. Novitski, A.V. Zlobin are with Fermi National Accelerator Laboratory, Batavia, IL 80510 USA.
M. Anerella, J. Cozzolino, R. Gupta, F. Kurian, and B. Yahia are with BNL, Upton, NY 11973-5000, USA.
L.D. Cooley is with the Applied Superconductivity Center, National High Magnetic Laboratory, Tallahassee, FL 32310, USA
E. Rochepault is with IRFU, CEA, Univers Paris-Saclay, Paris F-91191, France.
J. Stern is with TUFTS University, 419 Boston Ave, Medford, MA 02155, USA.

## II. Design Criteria and Conductor Parameters

The design criteria set as a goal of the conceptual design are given in Table I. The dipole has to generate a 20 T field of accelerator field quality with appropriate margin in a 50 mm clear bore. With respect to the criteria considered in [18], the target geometrical harmonics is reduced to <3 units. In addition, the maximum load-line fraction $I_{op}/I_{ss}$, i.e. the ratio between the operational current and the magnet current limit based on conductor properties (short sample current) is set to 87%, the same value adopted for the LHC dipoles [9] and similar to the 86% considered in the FCC design study [11]. Again, similarly to the FCC criteria, the maximum Von Mises stress allowed in the $Nb_3Sn$ coils is 180 MPa at 1.9 K; for the HTS conductor, a more conservative limit of 120 MPa has been assumed.

TABLE I
DESIGN CRITERIA ON MAGNET PARAMETERS

| Parameter | Unit | |
|---|---|---|
| Clear aperture | mm | 50 |
| Operational temperature | K | 1.9 |
| Operational bore field $B_{bore\_op}$ | T | 20 |
| Load-line fraction ($I_{op}/I_{ss}$) | % | 87 |
| Geometrical harmonics (20 T, $R_{ref}$=17 mm) | unit | <3 |
| Maximum $Nb_3Sn$ coil eq. stress (293 K) | MPa | 150 |
| Maximum $Nb_3Sn$ coil eq. stress (1.9 K) | MPa | 180 |
| Maximum HTS coil eq. stress (293K, 1.9 K) | MPa | 120 |
| Maximum hot spot temperature | K | 350 |

The two dashed lines in Fig. 1 depict the engineering current densities ($j_e = I_{strand}/A_{strand}$) used in the magnetic computations. For the $Nb_3Sn$ conductor, the curves correspond to a superconductor current density (virgin strand) of 3000 A/mm² at 12 T and 4.2 K (a level achieved within the US Conductor Development Program [20]), which, assuming a 1.1 Cu/Non-Cu ratio, results in a $j_e$ of 870 A/mm² at 16 T, 1.9 K, including 5% of cabling degradation. For the HTS conductor, we assumed a $j_e$ of 740 A/mm² at 1.9 K and 20 T. This current level was achieved in short samples of Bi2212 strands used in racetrack sub-scale coils [21].

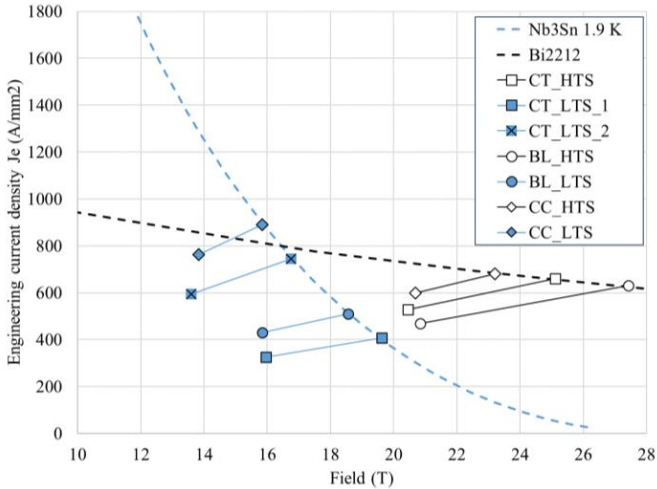

Fig. 1. Engineering current density ($j_e = I_{strand}/A_{strand}$) assumed in the computations for $Nb_3Sn$ and Bi2212 strands (dashed lines). Solid lines represent the load-lines defined by the operational and short sample currents (markers) for the cos-theta (CT), block (BL) and common-coil (CC) designs in the HTS and LTS coils.

## III. Sensitivity Analysis with Sector Coils

By simulating the superconducting coil as a 60° sector with a uniform overall current density ($j_o = I_{cable}/A_{ins\_cable}$) it is possible to carry out a sensitivity analysis where the key magnet parameters are investigated, as show in [22]-[23]. The magnetic numerical model (implemented in ANSYS 2D) assumes a 0.67 ratio between $j_o$ and $j_e$ (obtained by considering the $Nb_3Sn$ insulated cable for the MQXF project [24]) and a 250 mm thick iron yoke placed at 25 mm from the outer radius of the coil. In order to investigate the stress induced on the coil mid-plane by the azimuthal and radial electro-magnetic (e.m.) forces, the numerical mechanical model (implemented in ANSYS 2D) imposes an infinitely rigid structure all around the coil. The coil itself is also simulated with an infinity rigidity (to avoid bending effects) and with minimum shear modulus, in such a way that only the accumulation of e.m. forces on the mid-plane and on the outer radius are estimated. As output of the computations we focus on coil size, stresses and stored energies.

As a result of the slow and almost linear decrease in critical current as a function for the applied field observed in the HTS (see Fig. 1), the bore field increases almost linearly with the coil width, without exhibiting the "saturation" towards 10 T and 16 T observed in the Nb-Ti and $Nb_3Sn$ dipole magnets [23]. At a load-line fraction of 87%, a 20 T sector coil has a width of about 70 mm, compared to about 45 mm at 16 T (see Fig. 2).

The peak azimuthal and radial compressive stresses on the mid-plane due to the accumulation of the azimuthal and radial e.m. forces (see Fig. 3) reach -150 MPa with a bore field of 16 T and increase to more than -200 MPa at 20 T. This level of stress implies that stress management components have to be inserted in the coil design to reduce not only the azimuthal stress, as traditionally assumed, but also the radial stress, which appears to be the largest at 20 T and more dependent to the bore field.

With a value of 2.2 MJ/m, the 20T sector coil more than double the estimate of the stored energy for the 16 T (see Fig. 4). However, if the stored energy density over the insulated cable total area is considered, a value of 0.13 J/mm³ is obtained, still higher but more similar to the values computed or the FCC dipole magnets [25].

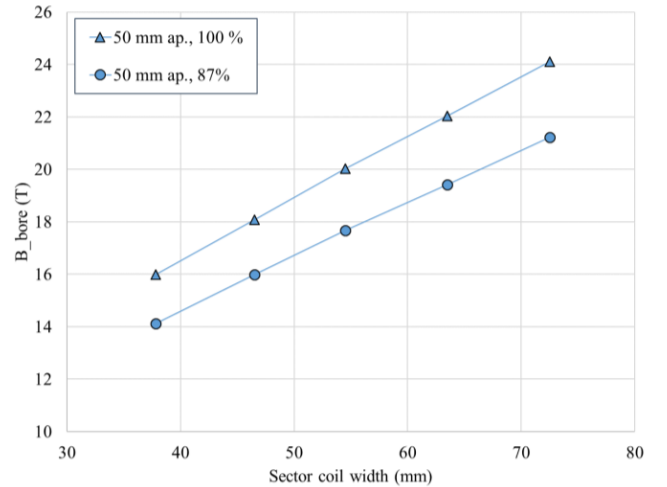

Fig. 2. Bore field vs coil width computed with a sector coil numerical model for an 87% and 100% load-line fraction $I_{op}/I_{ss}$.

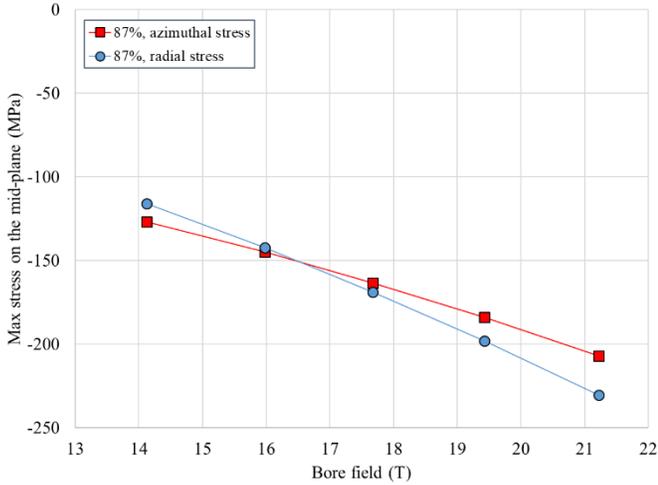

Fig. 3. Maximum azimuthal and radial stress on the mid-plane vs bore field computed with a sector coil numerical model for an 87% load-line fraction $I_{op}/I_{ss}$.

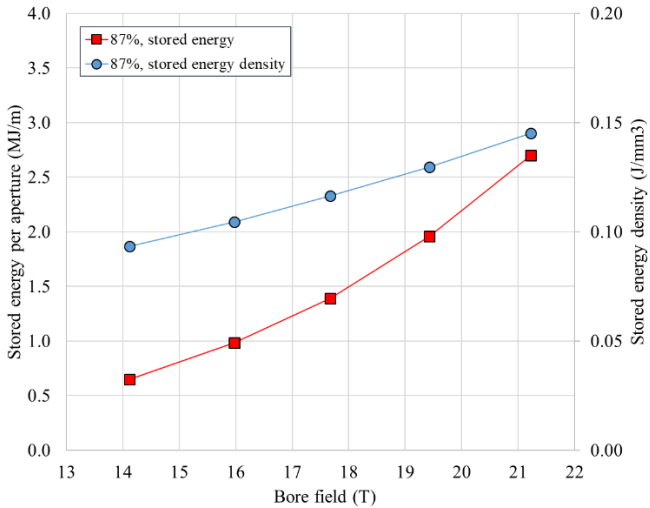

Fig. 4. Stored energy per aperture and stored energy density (considering the total insulated cable area) vs. bore field computed with a sector coil numerical model for an 87% load-line fraction $I_{op}/I_{ss}$.

## IV. CONCEPTUAL DESIGNS

In [18], 10 different designs were preliminary investigated to provide a first feedback on the general coils' size, load-line margin, and field quality. Starting from that analysis, we introduce in this paper the stress criteria provided in Table I. The results are described in the next sub-sections, where three designs are considered: a cos-theta (CT), a block (BL) and a common-coil (CC). The cable and magnet parameters of the three designs are summarized in Table II.

In terms of magnetic analysis, the strands diameters for both the $Nb_3Sn$ and HTS ranges from 0.85 to 1.15 mm, and the cable width from 13.3 mm to 24.4 mm. A cable compaction similar to the one of the MQXF cable [24] is assumed, again for both $Nb_3Sn$ and HTS cables. As for the sector coil analysis, a 250 mm thick iron is considered in the computations. The load-lines are shown in Fig.1, where the markers indicated the operational and shorts sample conditions.

As expected, meeting the coil stress criteria turned out to be the biggest challenge during the optimization of the coil design, since the high e.m. forces impose the use of stress management elements within the coil turns. The optimization was carried out to maintain the Von Mises stress below 120 MPa in the HTS [26], [27], and below 180 MPa in the $Nb_3Sn$, consistently to previous design studies [2], [11] and experimental studies [28], [29]. In addition, the following assumptions were set: 1) an elastic modulus of 25 GPa is associated to the coil turns and blocks; 2) the coil turns and blocks are surrounded by solid (i.e. "deformable") components made of stainless steel, bronze or Ti alloy (indicated in the following figure captions); 3) the coil turns and blocks are allowed to separate and slide with a 0.2 friction factor with respect to the stress management elements; 4) the surrounding iron yoke, not shown in the following cross-section figures is assumed to be infinitely rigid; 5) no pre-stress nor cool-down is applied. The mechanical analysis, whose results are described in the following sub-sections, is aimed exclusively at providing a first investigation on the level of stress interception and of the type of intercepting elements required to reduce the coil stresses produced only by the accumulation of the e.m. forces. It does not address the design of support structure, the pre-stress process, and cool-down conditions, which will be covered in the next phase of the conceptual design.

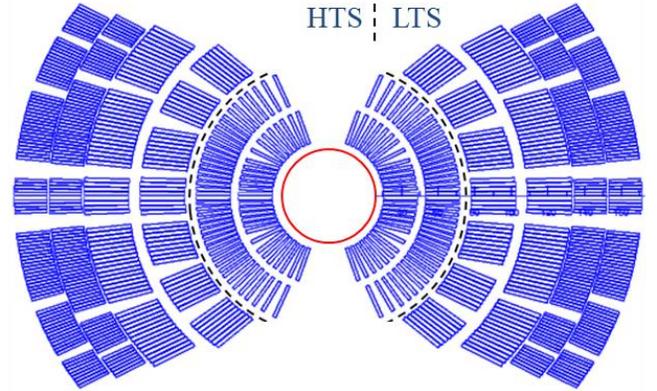

Fig. 5. Cross-section of the cos-theta (CT) design. The circle and the center of the coil aperture indicates the 50 mm clear aperture. The dashed line separates the HTS insert from and LTS outsert.

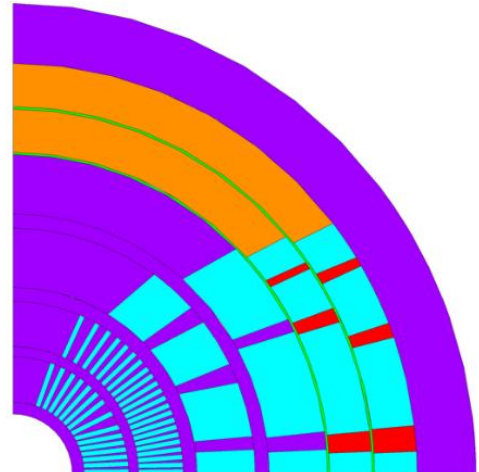

Fig. 6. Mechanical design of the cos-theta CT) design. The structural elements are assumed to be in stainless steel (purple), Ti alloy (orange) and Al-Br (red).

TABLE II
20 T Hybrid Magnet Parameters

| Parameter | Unit | CT$_{HTS}$ | CT$_{LTS\,I}$ | CT$_{LTS\,II}$ | BL$_{HTS}$ | BL$_{LTS}$ | CC$_{HTS}$ | CC$_{LTS}$ |
|---|---|---|---|---|---|---|---|---|
| Strand diameter | mm | 0.95 | 1.15 | 0.85 | 1.00 | 1.13 | 0.85 | 0.90 |
| N strands | | 36 | 40 | 40 | 28 | 24 | 40 | 28 |
| Cable width | mm | 18.590 | 24.380 | 17.730 | 14.700 | 14.700 | 18.350 | 13.300 |
| Cable mid-thickness | mm | 1.705 | 2.085 | 1.515 | 1.800 | 2.030 | 1.520 | 1.600 |
| Insulation thickness | mm | 0.150 | 0.150 | 0.150 | 0.150 | 0.150 | 0.150 | 0.150 |
| Clear aperture | mm | | 50 | | | 50 | | 50 |
| Coil aperture* | mm | | 60 | | | 70 | | 50 |
| N turns per quadrant | | 37 | 64 | 95 | 56 | 210 | 42 | 105 |
| Area ins. cable per quadrant | mm$^2$ | 1401 | 3767 | 3109 | 1764 | 7340 | 1426 | 2713 |
| Current_op | kA | | 13.5 | | | 10.3 | | 13.6 |
| B_bore_op | T | | 20.0 | | | 20.0 | | 19.9 |
| B_peak_op | T | 20.5 | 16.0 | 13.6 | 20.84 | 15.85 | 20.7 | 13.8 |
| J$_e$ _op | A/mm$^2$ | 529 | 325 | 595 | 470 | 429 | 599 | 763 |
| Magnet current_ss | kA | | 16.8 | | | 12.3 | | 15.4 |
| B_bore_ss | T | | 24.6 | | | 23.5 | | 22.4 |
| Load-line ratio | % | 80 | 80 | 80 | 75 | 84 | 88 | 86 |

* Given by the inner radius of the innermost cable on the mid-plane.

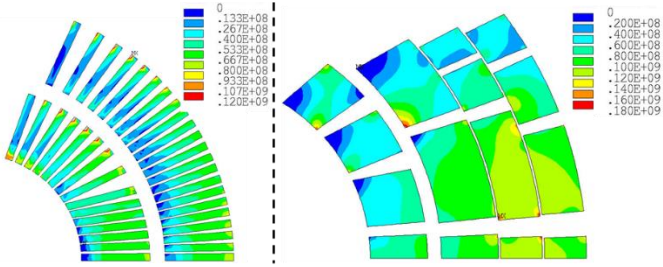

Fig. 7. Von Mises stress (Pa) in the conductor under the action of e.m. forces: HTS inserts (left) and LTS outsert (right).

## A. Cos-theta (CT) Design

The cross-section of the cos-theta design, analyzed in details in [30], is shown in Fig. 5, where the central red circle represents the 50 mm clear aperture and the dashed lines indicate the separation between the HTS insert and the LTS outsert. The layout is characterized by three double-layer coils wound with a continuous cable unit length. This option prevents the use of internal splices, as in most of the cos-theta Nb$_3$Sn coils fabricated so far, with the exception of the CERN-ELIN and UT-CERN dipole magnets [12]. In the innermost two layers HTS cable turns are wound into individual slots in the coil support structure, as in a canted cos-theta (CCT) design [31]-[33]. In the two central layers, groups of turns (turn blocks) are wound in the coil structure groves, as it is done in the Stress Management cos-theta (SMCT) design [34]-[36]. Finally, the two outermost layers can be defined a traditional cos-theta coil with turn blocks separated by spacers [37], [38].

The cable width ranges from 17.7 mm in layers 5-6 to 24.4 mm in layer 3-4. The use of wider cable in layer 3-4 compared to layer 1-2 is aimed at minimizing the size of the HTS coils by increasing the size of the LTS ones, a design choice inspired to the "anti-grading" sector coils shown in [18].

In operational conditions with a bore field of 20 T, the calculated geometrical harmonics are within 3 units, the conductor peak field is 20.5 T in the HTS and 16.0 T in the LTS, and the corresponding load-line ratio is 80% in all coils.

The use of three different cos-theta coil designs is exclusively related to the outputs of the mechanical analysis. In fact, the combined effect of deformation induced by the large e.m. forces and of the low stress limit of 120 MPa assumed for the HTS coils could be overcome only by implementing a high level of stress interception (see Fig. 6).

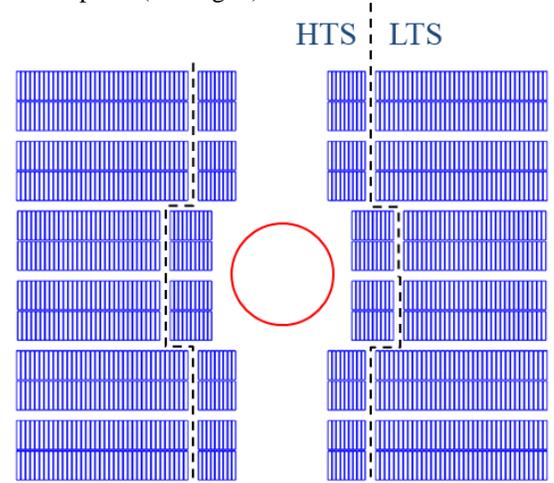

Fig. 8. Cross-section of the block (BL) design. The circle and the center of the coil aperture indicates the 50 mm clear aperture. The dashed line separates the HTS insert from and LTS outsert.

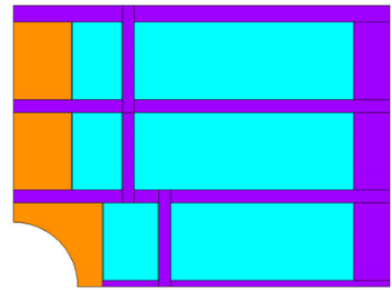

Fig. 9. Mechanical design of the block (BL) design. All the structure elements are assumed to be in stainless steel (purple) and Ti alloy (orange).

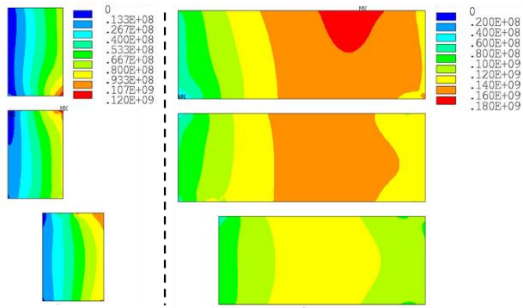

Fig. 10. Von Mises stress (Pa) in the conductor under the action of e.m. forces: HTS inserts (left) and LTS outsert (right).

This is the case in the CCT-like layer 1-2, where each turn is separated by ribs. The ribs have a minimum thickness of 0.4 mm and are connected to a 5 mm spar (or mandrel). In the layer 3-4, a lower level of stress interception, magnetically more efficient, was adopted to maintain the $Nb_3Sn$ coil stress level below 180 MPa, where coil blocks (not the individual turns) are separated by ribs, following the SMCT design. Finally, no stress management elements were used in layer 5-6. As can be seen in Fig. 7, both HTS and LTS coils have Von Mises stress under the limits established by the design criteria, except for small corner effects (gray areas in Fig. 7, left) in the HTS turns of layer 2.

### B. Block (BL) Design

The block design, shown in Fig. 8 and analyzed in [39], features also three double-layers coils, all composed by narrow HTS inner blocks and wide LTS outer blocks. As for the CT option, no internal splices are assumed. The overall design follows the main characteristics of the HD2 [40] and FRESCA2 [41] designs and of other conceptual designs [42], [43], with blocks aligned in the outer edge. The cable width is 14.7 mm for both HTS and LTS coils, but, similarly to the CT design, with a higher thickness in the LTS. The design meets the field quality requirements, and with a bore field of 20 T it operates at a load-line ratio of 75% in the HTS and 84% in the LTS. Also, both the HTS and the LTS coil area are similar to the CT design.

The mechanical design (see Fig. 9) is characterized by a 10 mm thick internal support (winding pole) which brings the coil aperture to 70 mm. A similar support was implemented in both HD2 and FRESCA2. In addition, the coils are vertically separated by horizontal plates, which provide vertical stress management, and by vertical ribs, which separate the HTS and LTS blocks and provide horizontal stress management. In particular, the ribs transfer the horizontal e.m. force to the horizontal plates, in a way that maintains the coil stress within the limits in both the LTS and HTS. Horizontal plates aimed at intercepting the vertical forces were included in the design of the Test Facility Dipole [44]. The most challenging aspect of the optimization consisted in minimizing the bending of the ribs, which could generate extremely high stress in the corners of the coil blocks. A solution was found by including gaps (or clearances) of 0.200 to 0.300 mm between the ribs and the plates. Under these conditions, only an initial small fraction of the e.m. force is transferred from the HTS blocks to the LTS blocks. And once the ribs get in contact with the plates, the force is transmitted to the latter, and the ribs bending is minimized. The results of the mechanical analysis are shown in Fig. 10, with all the stress within the design criteria.

As a last general consideration regarding the block design, it is important to point out that at the moment no block coil has been fabricated with different cables sizes (grading) or different superconducting materials (hybrid). Therefore, inserting an HTS block coil inside an LTS block coil appears to be the biggest design and fabrication challenge for this option. Possible fabrication and assembly solutions for this issue are provided in [39].

### C. Common-Coil (CC) Design

The common-coil design (CC) is characterized by large race-track coils that cover both apertures [45]-[47]. In Fig. 11, the coil cross-section of one aperture is shown. Unlike the CT and BL designs, the coil aperture and the clear aperture are identical, so no internal support is considered, similarly to [46]. The HTS part is composed by two blocks (per quadrants) close to the aperture, each with two turns, and by a single-layer large racetrack coil. All HTS blocks are wound with an 18.4 mm wide cable.

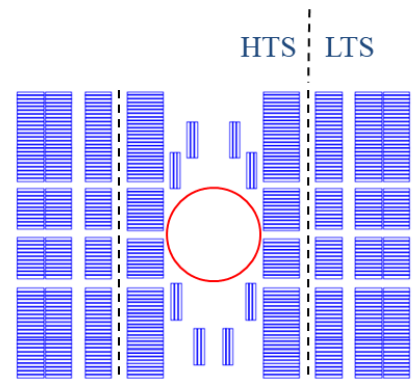

Fig. 11. Cross-section (one aperture) of the common-coil (BL) design. The circle and the center of the coil aperture indicates the 50 mm clear aperture. The dashed line separates the HTS insert from and LTS outsert.

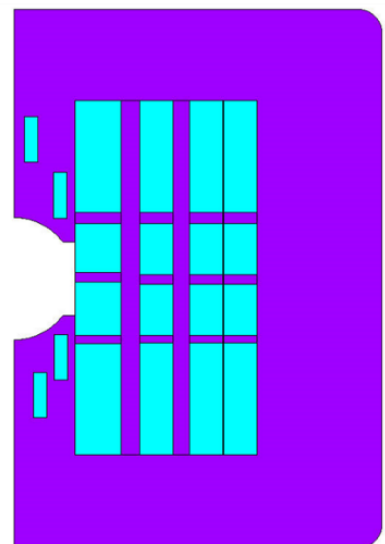

Fig. 12. Mechanical design of the common-coil (CC) design. All the structure elements are assumed to be in stainless steel.

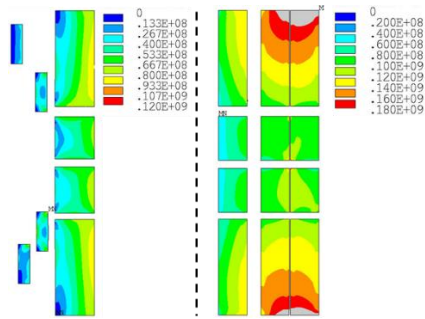

Fig. 13. Von Mises stress (Pa) in the conductor under the action of e.m. forces: HTS inserts (left) and LTS outsert (right).

The two-turns blocks close to the aperture, often referred as "pole coils", have the main function of correcting the field quality, and they were included also in previous design studies [6], [45]-[47]. Since pole coils require some sort of hard-way bend of the cable to clear the path of the bore tube, they represent a departure from the typical common-coil advantage of using simple racetrack coils. However, the bending radius remains significantly larger compared to the CT design.

and, not being ever implemented in previous CC magnets, they constitute a design and fabrication challenge. In fact, since some sort of hard-way bend of the cable is required to clear the path of the bore tube, they represent a departure from the typical common-coil advantage of using simple racetrack coils.

The $Nb_3Sn$ part of the coil is composed by three layers, all using the same 13.3 mm wide cable. Unlike the CT and BL design, a single layer coil can be easily connected to another single layer coil, thanks to the wide central winding pole which can provide enough real estate for the splicing operation. Therefore, double-layer coils were not imposed to the CC design, as was done for the previous two designs. Another important characteristic of the CC lay-out is that the vertical dimensions of the layers can be easily fine-tuned by simply stacking or removing turns. This possibility is not available for example in the BL design, were the vertical dimensions are defined by layers with a given cable width. These two advantages of the CC design (single layer coils and vertical tunability of blocks' size) provide an additional flexibility in the optimization of the coil shape compared to the CT and BL designs.

The CC design has all geometric harmonics below 3 units, and load-line ratio is is within 1 % of the limits set in the criteria, i.e. 88% in the HTS and 86% in the LTS.

Stress management in the CC design is obtain again by vertical plates and horizontal ribs (see Fig. 12). The vertical plates are allowed to slide with respect to the external collars. Similarly, the ribs are allowed to slide with respect to the plates. As a result, no vertical stress management is provided, and only the horizontal forces are intercepted, in this case by the vertical plates supported by the horizontal ribs. With this mechanical design, the stress in the HTS blocks is maintained within 120 MPa. However, stresses higher than 180 MPa can be seen in the top part of the LTS coils (see Fig. 13).

The total area of the HTS block is similar to the one of the CT and BL designs, but a significant lower area for the LTS is observed in the CC. However, it is important to point out that the CC has a lower coil aperture, a lower load-line margin, and still a higher conductor peak stress in the LTS compared to the CT and BL designs.

## V. CONCLUSIONS

We presented in this paper the conceptual design of a dipole magnet with an operational field of 20 T, generated by a hybrid coil made with both HTS and LTS ($Nb_3Sn$) superconducting materials. The analysis included both a magnetic study, focused on bore field, load-line ratio and field quality, and a mechanical study, aimed at keeping the Von Mises stress below 180 (120) MPa in the LTS (HTS) conductor. An initial analytical/numerical study using sector coils indicated that in a 20 T dipole magnet, 1) the coil has to be about 70 mm wide, 2) both radial and azimuthal stress in the coil induced by the accumulation of the e.m. forces are above 200 MPa, and 3) the stored energy densities in the insulated cables are of about 0.13 J/mm$^3$. Three were the design options analyzed, all with stress management elements: 1) a cos-theta design, including CCT-like SMCT traditional cos-theta two-layer coils, 2) a block-type design, and 3) a common-coil design. All layouts meet the bore field, margin, and field quality requirements. In terms of conductor quantity, the designs have similar HTS conductor area, while a lower LTS area is obtained in the common-coil. The mechanical analysis showed that the cos-theta option requires individual turn support in the HTS layers and coil blocks support in the inner LTS layers to reduce the coil peak stress. Also, in both the block and common coil designs, a series of plates and ribs are necessary to intercept the e.m. forces and to keep the accumulated stress within the limits.